\documentclass[twocolumn,preprintnumbers,amsmath,superscriptaddress,amssymb,aps]{revtex4-1}

\usepackage{graphicx}
\usepackage{dcolumn}
\usepackage{bm}
\usepackage{amsmath}
\usepackage{amssymb}
\usepackage{graphicx}
\usepackage{indentfirst}
\usepackage{booktabs}
\usepackage{multirow}
\usepackage{colortbl}
\usepackage{float}

\selectfont

\begin{document}
\title{Sliding Ferroelectricity Induced and Switched Altermagnetism in GaSe-VPSe$_3$-GaSe Sandwiched Heterostructure with Strong Magnetoelectric Effect}
\author{Pengqiang Dong}
\address{State Key Laboratory for Mechanical Behavior of Materials, School of Materials Science and Engineering, Xi'an Jiaotong University, Xi'an, Shaanxi, 710049, People's Republic of China}
\address{Shaanxi Yanyi Titanium Industry Company Limited, Baoji, Shaanxi, 721001, People's Republic of China}
\author{Hanbo Sun}
\address{State Key Laboratory for Mechanical Behavior of Materials, School of Materials Science and Engineering, Xi'an Jiaotong University, Xi'an, Shaanxi, 710049, People's Republic of China}
\address{Shaanxi Yanyi Titanium Industry Company Limited, Baoji, Shaanxi, 721001, People's Republic of China}
\author{Chao Wu}
\address{State Key Laboratory for Mechanical Behavior of Materials, School of Materials Science and Engineering, Xi'an Jiaotong University, Xi'an, Shaanxi, 710049, People's Republic of China}
\address{Shaanxi Yanyi Titanium Industry Company Limited, Baoji, Shaanxi, 721001, People's Republic of China}
\author{Ping Li}
\email{pli@xjtu.edu.cn}
\address{State Key Laboratory for Mechanical Behavior of Materials, School of Materials Science and Engineering, Xi'an Jiaotong University, Xi'an, Shaanxi, 710049, People's Republic of China}
\address{Shaanxi Yanyi Titanium Industry Company Limited, Baoji, Shaanxi, 721001, People's Republic of China}
\address{National Laboratory of Solid State Microstructures, Nanjing University, Nanjing, 210093, People's Republic of China}
\address{State Key Laboratory of Silicon and Advanced Semiconductor Materials, Zhejiang University, Hangzhou, Zhejiang, 310027, People's Republic of China}

\date{\today}

\begin{abstract}
Magnetoelectric coupling is vital for exploring fundamental science and driving the development of high-density memory and energy-efficient spintronic devices. Altermagnets, which merge the benefits of ferromagnets and antiferromagnets, pave the way for unprecedented magnetoelectric coupling effects. However, the spin splitting in altermagnets is robustly protected by spin space group symmetry, posing a significant challenge for external manipulation. Here, we propose to utilize the coupling between the layer degree of freedom and the altermagnet to achieve an altermagnetic multiferroic with strong magnetoelectric coupling. In the GaSe-VPSe$_3$-GaSe sandwiched structure, the magnetic order can be switched between altermagnetic and conventional antiferromagnetic by controllably breaking and restoring the combined spatial inversion and time-reversal symmetry using sliding ferroelectricity. Moreover, our systematic investigation of all pathways revealed that the transition from a ferroelectric CB stacking, through an antiferroelectric CC stacking, to a ferroelectric BC stacking is the most favorable, with an energy barrier of only 50.13 meV/f.u.. More importantly, we reveal that the microscopic mechanism of the magnetic phase transition stems from the interlayer covalent bonding of Se-Se or Se-P atomic pairs at the interface. Our findings unveil a new form of magnetoelectric coupling and lay the groundwork for designing miniature information processing and multiferroic memory devices based on altermagnetism.
\end{abstract}

\maketitle
\section{Introduction}
The recent discovery of altermagnetism, a new type of collinear magnetic state, extends the classification of magnetic materials beyond the conventional descriptions of ferromagnetism and antiferromagnetism \cite{1,2,3,4,5,6,7,8}. Altermagnetism exhibits unique characteristics, which is distinguished from ferromagnets (FM) that possess a net magnetization and antiferromagnets (AFM) that display spin degeneracy. In altermagnets, an alternating non-relativistic spin splitting (NRSS) is exhibited in momentum space in a manner dictated by the lattice symmetry and broken time-reversal symmetry ($\emph{T}$), yet the net magnetization integrates to zero over the Brillouin zone \cite{9,10}. The spin-dependent Fermi surfaces of altermagnet exhibits the $\emph{d}$-, $\emph{g}$-, or $\emph{i}$-wave symmetry in momentum space \cite{1,8,9,10}. These unique magnetic and electronic properties underlie a range of multifunctional applications. The unconventional spin polarization of altermagnets enables the production and manipulation of spin currents, making them ideal materials for low-power and high-performance spintronic devices \cite{11,12,13,14,15,16,17}. Moreover, the significant anomalous Hall effect \cite{18,19}, the giant and tunneling magnetoresistance effect \cite{20,21} can be used for the information processing and storage. Despite significant progress in altermagnets for spintronics, the realization of magnetoelectric devices remains a challenge. Combining altermagnetism with ferroelectricity offers a pathway to multifunctional materials that could enhance the functionality of devices such as nonvolatile memories.

\begin{figure*}[htb]
\begin{center}
\includegraphics[angle=0,width=1.0\linewidth]{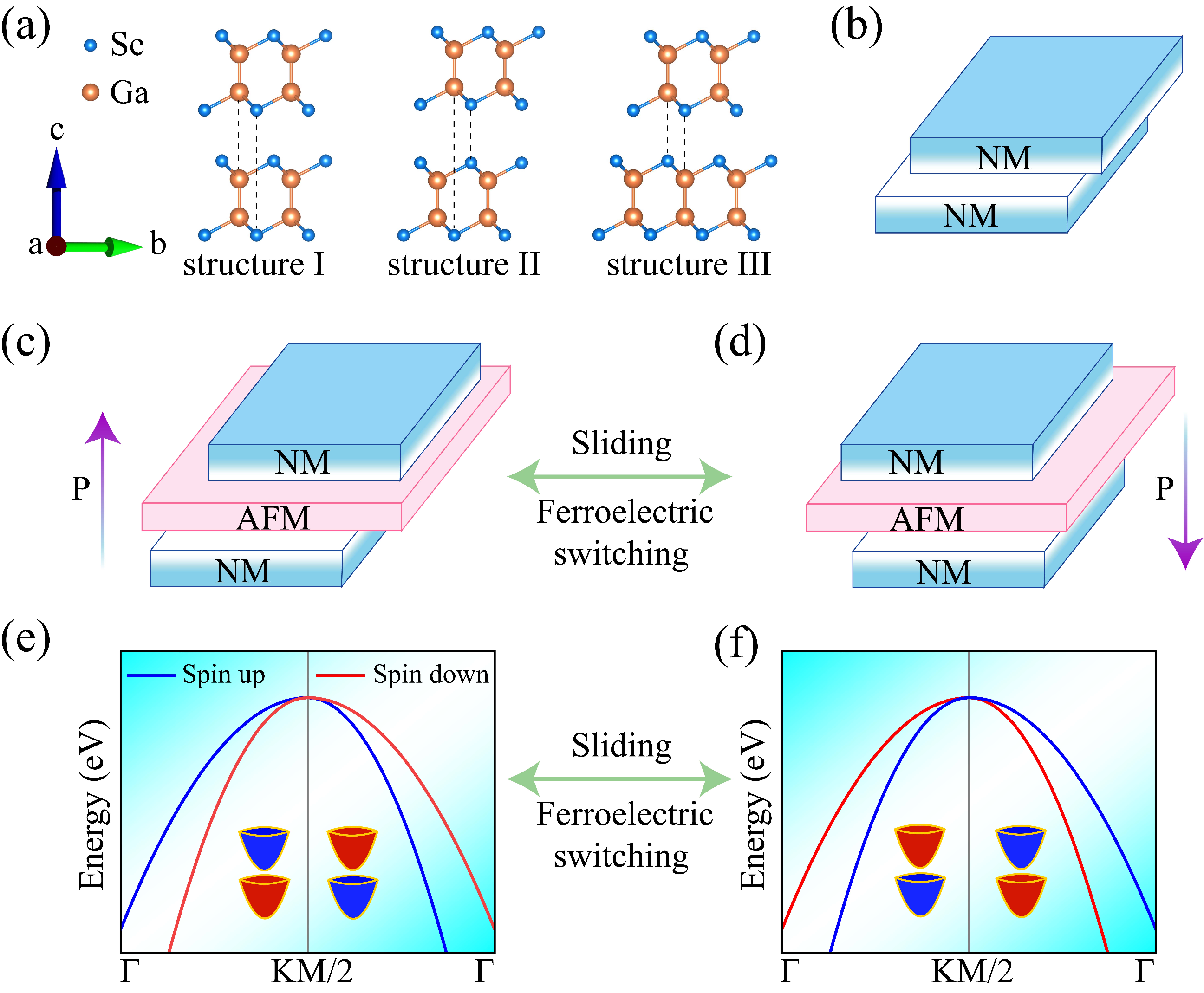}
\caption{
(a) The side view of bilayer GaSe crystal structures. The blue and orange balls represents Se and Ga atoms, respectively. (b) The schematic diagram of bilayer GaSe stacking. (c, d) The schematic diagram of the constructed sandwich structure from bilayer GaSe. The sandwich structure is (c) polarization upward and (d) polarization downward. The polarization direction can be manipulated by interlayer sliding. (e, f) show band structures diagram of (c, d) sandwich structures. The altermagnetic effect can also be tuned via ferroelectric switching.
}
\end{center}
\end{figure*}

To promote the practical application, it is important to transform the altermagnets into multiferroic materials by sliding ferroelectricity. Sliding ferroelectricity breaks the traditional ferroelectric critical size effect and perfect compatibility with two-dimensional (2D) materials \cite{22,23,24,25,26}. In generally, out-of-plane polarization is induced by non-polar materials with central symmetry through interlayer relative sliding or twisting, thereby breaking the spatial inversion symmetry ($\emph{P}$) \cite{27,28,29,30}. Its polarization reversal mode ingeniously takes advantage of the weak interlayer coupling in van der Waals (vdW) 2D material layers, avoiding the traditional process of domain wall movement for polarization reversal \cite{31,32}. Moreover, sliding ferroelectricity can effectively tune various quantum phenomena, such as, inducing magnetic phase transition in bilayer CrI$_3$ and CrBr$_3$ \cite{33,34}; realizing hybrid-order topological phase transitions \cite{35}; and tuning valley polarization and layer-polarized anomalous Hall effect \cite{36,37,38,39}. Therefore, they have developed highly performing multiferroic materials.

Multiferroic materials have garnered considerable attention due to their capability to manipulate magnetic behaviors via ferroelectric polarization and vice versa, paving the way for novel devices in data storage, sensing, and spintronic technologies \cite{40,41,42,43}. This originates from the fact that multiferroic materials contain more than one type of ferroic order (i.e., ferromagnetic/antiferromagnetic, ferroelectric/antiferroelectric, ferroelastic/antiferroelatic, ferrovalley/antiferrovalley) \cite{44,45,46,47,48,49}. The emergence of altermagnets has led to extensive research on altermagnetic multiferroics. For example, Liu and Zhou et al. proposed that ferroelectricity can effectively switch the altermagnetic effect, and this was confirmed in [C(NH$_2$)$_3$]Cr(HCOO)$_3$ and VOI$_2$ \cite{50,51,52}. Moreover, based on the unique symmetry of the altermagnetism, the type-III multiferroic is proposed \cite{53}. However, the mutual exclusivity of magnetism and ferroelectricity leads to a relatively weak magnetoelectric coupling. If the sliding ferroelectric and the altermagnets can be combined together, it will offers a promising and practical approach for creating a novel category of multiferroics that exhibit strong magnetoelectric coupling.

In this work, we propose a new strategy for designing 2D altermagnetic multiferroics, employing sliding ferroelectric to modulate lattice symmetry and consequently initiate unprecedented altermagnetic phase transitions. Here we construct a sandwiched structure GaSe-VPSe$_3$-GaSe, through symmetry analysis and density functional theory (DFT) calculations, we confirm that the sliding ferroelectric can induce the system to transform from the a traditional spin degenerate AFM to an altermagnet with spin splitting. This symmetry configuration dictates two distinct sliding methods between the upper and lower sublattices: sequential sliding (upper layer first, followed by lower) and inverted sliding (lower layer first, followed by upper). Through a systematic investigation, we find three sliding paths, which the sliding barrier for the CB-stacking to CC-stacking to BC-stacking path is only 50.13 meV/f.u.. Meanwhile, the NRSS of the BC and CB stacking are as high as 51.41 meV. Then, an electric field can effectively tune the magnitude of NRSS. In addition, we uncover the microscopic mechanism of the magnetic phase transition by the symmetry analysis and interlayer charge transfer. Our research not only expandeds the definition of multiferroic materials, but also provides a way to realize strong magnetoelectric coupling mediated through lattice symmetry from a fundamentally novel perspective.

\section{Methods}
Based on the framework of density functional theory (DFT), we employed the Vienna $Ab$ $initio$ Simulation Package (VASP) to investigate the stability, magnetic properties, sliding ferroelectricity, and electronic properties of the sandwich structure\cite{54,55,56}. The treatment of the exchange-correlation energy uses the generalized gradient approximation (GGA) with the Perdew-Burke-Ernzerhof (PBE) functional\cite{57}. The kinetic energy cutoff for the plane-wave basis is set to 500 eV. The calculations employs force and energy convergence thresholds of -0.005 eV/$\rm \AA$ and 10$^{-6}$ eV, respectively. A vacuum spacing of 40 $\rm \AA$ is used along the c-axis to prevent interactions between periodic images of the sheet. The k-point grid employs a $12\times 12\times 1$ on the $\Gamma$-centered. Moreover, we applied the DFT-D3 method with zero-damping to account for the vdW interactions in the sandwich structure \cite{58}. The GGA+U method, with an effective U parameter (U$_{eff}$ = U - J) of 3 eV, is employed to treat the strongly correlated 3d electrons of V \cite{59}.

\section{RESULTS AND DISCUSSION }	
\subsection{Model and Symmetry}
First, we define the stacking configuration of bilayer GaSe. As shown in Fig. 1(a), we name the three typical stacking structures as structure I, structure II, and structure III, respectively. To present the GaSe-VPSe$_3$-GaSe sandwich structure more intuitively, as shown in Fig. 1(b), we first provide a schematic diagram of the bilayer GaSe structure. Then, we exhibit two polarization states of the sandwich structure schematic diagram. As illustrated in Fig. 1(c, d), the two ferroelectric polarization states can be transformed into each other through interlayer sliding. More interestingly, when the ferroelectric polarization is upward, the system is an altermagnetic state. As shown in Fig. 1(e), the spin up and spin down bands exhibit spin splitting, regarding the KM/2 point symmetry. When the sliding switches the ferroelectric polarization downward, as shown in Fig. 1(d), the switching of altermagnetic spin splitting occurred simultaneously with the switching process (see Fig. 1(f)). The strong correlation between the altermagnetic spin degree of freedom and the electric polarization suggests a distinct magnetoelectric coupling mechanism.

To have a comprehensive understanding of the underlying mechanism, we first analyze the symmetry of the sandwich structure. For monolayer AFM VPSe$_3$, it breaks the $\emph{T}$ symmetry, but has the $\emph{P}$ and $\emph{PT}$ combined symmetry. Moreover, the opposite spin sublattices are linked through a symmetry operator that combines rotation and mirror operations, defined as [C$_2$$\|$M]. The [C$_2$$\|$M] symmetry guarantees the energy relation $\emph{E}$(s, $\textbf{k}$) = [C$_2$$\|$M]$\emph{E}$(s, $\textbf{k}$) = $\emph{E}$(-s, $\emph{M}$$\textbf{k}$), which lays an important foundation for the realization of altermagnets. The $\textbf{k}$, $\emph{s}$, and $\emph{E}$(s, $\textbf{k}$) denote the momentum, spin, and spin-momentum-dependent bands, respectively. For the sandwiched structure of GaSe-VPSe$_3$-GaSe, when the system is in the paraelectric phase, $\emph{PT}$ symmetry will ensure spin degeneracy throughout the entire Brillouin zone. However, when it shifts to the ferroelectric phase, it will break the $\emph{PT}$ symmetry while preserving [C$_2$$\|$M] symmetry. At this stage, the system undergoes magnetic phase transition from a conventional AFM state to an altermagnetic state, with the manifestation of spin splitting demonstrated in Fig. 1 (e, f). Although the ferroelectric polarization breaks the $\emph{PT}$ symmetry, it now serves to couple the two states (denoted E$_+$ and E$_-$) of an extrinsic altermagnet under positive and negative ferroelectric polarization (P$\uparrow$ and P$\downarrow$). Therefore, the sign of spin splitting will alter after switching the direction of polarization due to the $\emph{E$_+$}$(s, $\textbf{k}$) = $\emph{PT}$$\emph{E$_+$}$(s, $\textbf{k}$) = $\emph{E$_-$}$(-s, $\textbf{k}$), as shown in Fig. 1(e, f).

\begin{figure*}[htb]
\begin{center}
\includegraphics[angle=0,width=1.0\linewidth]{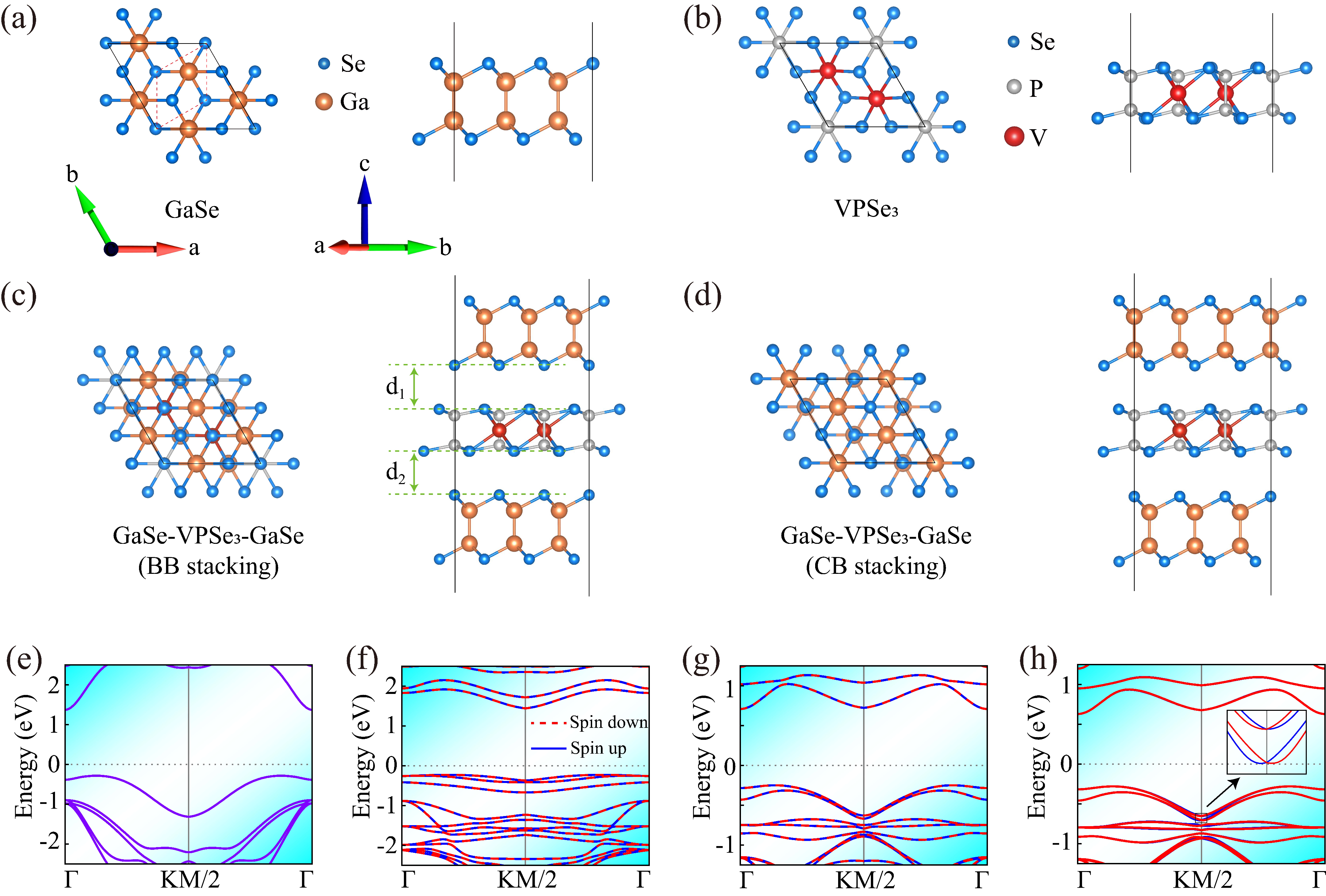}
\caption{
(a) The top and side views of monolayer GaSe. The blue and orange balls represents Se and Ga atoms, respectively. (b) Crystal structures of monolayer VPSe$_3$ from top and side views. The blue, silver, and red balls represents Se, P, and V atoms, respectively. (c, d) The top and side views of GaSe-VPSe$_3$-GaSe sandwiched structure for (c) BB stacking and (d) CB stacking. The layer spacing is marked as d$_1$ and d$_2$. (e-h) The band structures without SOC of (e) monolayer GaSe, (f) monolayer VPSe$_3$, (g) BB stacking GaSe-VPSe$_3$-GaSe, and (h) CB stacking GaSe-VPSe$_3$-GaSe. The blue line and red line (red dotted line) denote spin up and spin down bands, respectively.
}
\end{center}
\end{figure*}

Then, based on the symmetry analysis, we need to verify the proposed mechanism in the GaSe-VPSe$_3$-GaSe sandwiched structure. Primarily, we investigate the structure and band structures of monolayer GaSe and VPSe$_3$. Fig. 2(a, b) shows the crystal structure of monolayer GaSe and VPSe$_3$, which the space group and point group are P$\bar{6}$m2, D$_{3h}$ and P$\bar{3}$1m, D$_{3d}$, respectively. The optimized lattice constant is 3.83 $\rm \AA$ of monolayer GaSe and 6.24 $\rm \AA$ of monolayer VPSe$_3$. Monolayer GaSe is nonpolar semiconductor with an indirect band gap of 1.65 eV (see Fig. 2(e) and Fig. S1(a)), while monolayer VPSe$_3$ is a traditional AFM semiconductor with a band gap of 1.45 eV (see Fig. 2(f) and Fig. S1(b)). Considering their lattice mismatch rate, we use 1 $\times$ 1 VPSe$_3$ to match $\sqrt{3}$ $\times$ $\sqrt{3}$ GaSe with the lattice mismatch of 5.94 $\%$. To comprehensively explore the electronic properties of the GaSe-VPSe$_3$-GaSe sandwiched structure, we consider nine typical stacking configurations, namely, AA, AB, BA, BB, BC, CB, CC, AC, and CA stackings, as shown in Fig. 2(c, d) and Fig. S2. In our naming convention, the first and second letters denote the relationship between the V atom and the upper and lower neighboring GaSe hexagonal rings, respectively. The labels A, B, C denote that the V atom is positioned directly above the hollow, top-Se, and top-Ga positions of the GaSe hexagon, respectively. As listed in Table SI, the CC stacking is the most stable GaSe-VPSe$_3$-GaSe sandwiched structure. For AA, BB, CC stackings with the $\emph{PT}$ symmetry, as shown in Fig. 2(g) and Fig. S3(a, d, g), the spin up and spin down bands are completely degenerate throughout the Brillouin zone. On the contrary, when the $\emph{PT}$ symmetry is broken, as shown in Fig. S3(b, c, e, f, h, i), the typical spin splitting feature of an altermagnet will exhibit along the $\Gamma$-KM/2-$\Gamma$ path. In addition, Fig. S4 shows an enlarged view of the $\Gamma$-KM/2-$\Gamma$ path for the $\emph{PT}$ symmetry broken configuration. This demonstrates that our proposed physical mechanism is feasible.

\subsection{Sliding Path and The Most Stable Sandwich Configuration}
Through symmetry analysis, it is known that among the nine typical stacking sequences, AB, BA, BC, CB, AC, and CA stackings exhibits out-of-plane ferroelectric polarization. As shown in Fig. S5, a finite difference in the average electrostatic potential exists in the ferroelectric configuration, while it vanishes in the antiferroelectric configuration. This provides a clear confirmation that these stackings have out-of-plane ferroelectric polarization. It is noteworthy that AB, BC, and AC stackings sequences transform into BA, CB, and CA stackings, respectively, under a 180$^\circ$ rotation about the xy-plane. Since both the lower and upper GaSe layers can slide relative to the VPSe$_3$, there exist two distinct slip modes within the GaSe-VPSe$_3$-GaSe sandwiched structure. One involves first sliding the upper GaSe layer and then the lower GaSe layer, while the other follows the reverse sequence. We designate these as path A and path B. Therefore, we comprehensively investigate the pathway for the conversion of polarization from AB, CB, and AC stackings with upward polarization to BA, BC, and CA stackings with downward polarization, as shown in Fig. 3. Here, it should be noted that the simultaneous sliding of the two layers requires overcoming a higher energy barrier compared to monolayer GaSe sliding (see Fig. S6). Hence, it is excluded from the following discussion.

\begin{figure*}[htb]
\begin{center}
\includegraphics[angle=0,width=1.0\linewidth]{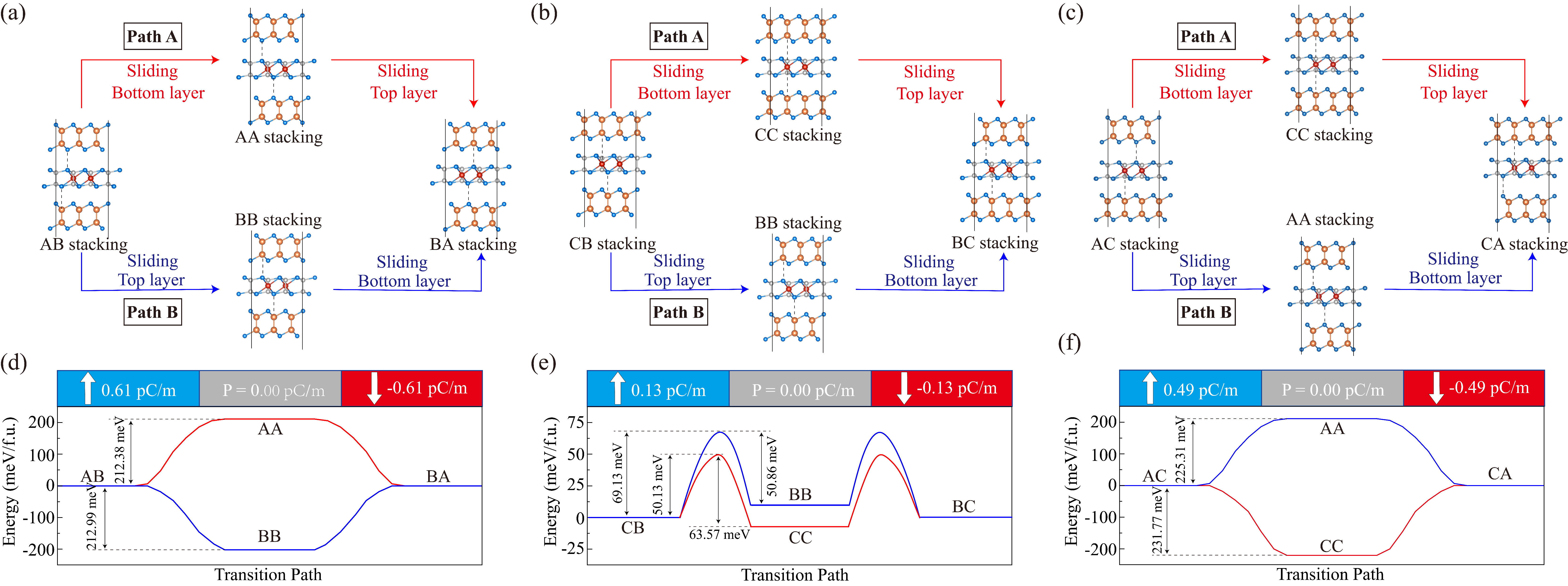}
\caption{
(a-c) Two ferroelectric transition paths of GaSe-VPSe$_3$-GaSe sandwiched structure formed (a) from AB stacking to BA stacking, (b) from CB stacking to BC stacking, and (c) from CA stacking to AC stacking by sliding sequential order. (d-f) Sliding energy barriers (d) from AB stacking to BA stacking, (e) from CB stacking to BC stacking, and (f) from CA stacking to AC stacking, respectively.
}
\end{center}
\end{figure*}

As shown in Fig. 3, we find that the three types of stacking conversion can be classified into two categories. One type is the sliding from AB stacking to BA stacking and AC stacking to CA stacking, exhibiting a significant change in energy barriers from positive to negative values (see Fig. 3(a, d, c, f)). In this case, when the sliding path passes through the AA stacking, the energy barrier is as high as 212.38 meV/f.u. and 225.31 meV/f.u.. However, there is no energy barrier for the other path. This originates from the significant repulsive interaction between the Se-Se atomic pairs at the two interfaces of the AA stacking, which substantially increases its total energy. The interlayer spacing in Table SII clearly demonstrates this point. Due to the repulsive interaction between Se-Se atomic pairs, the interlayer spacing of the AA stacking has a maximum value of 3.94 $\rm \AA$. The eventual result is that the AA stacking exhibits the highest total energy (see Table SI). In the AB, BA, AC, and CA stackings, an interface of configuration A (i.e., a Se-Se atomic pair) still exists. As the slip path passes through the BB and CC stackings, the sliding barrier will naturally vanish. Its essence is that the repulsive force caused by the Se-Se atomic pairs disappears, and the total energy decreases. It should be noted that the ferroelectric polarizations for the AB, BA, AC, and CA stackings with broken $\emph{P}$ symmetry are 0.61 pC/m, -0.61 pC/m, 0.49 pC/m, and -0.49 pC/m (see Table SIII), respectively. This may originate from the charge transfer between the nonequivalent Se-P atomic pairs at the two interfaces.

Another type is the sliding from CB to BC stacking, as shown in Fig. 3(b, e), which shows a significant reduction in the energy barrier. The energy barrier for path A is only 50.13 meV/f.u., which is 19.00 meV/f.u. lower than that for path B. This indicates that path A is more favorable for the ferroelectric polarization switching process. In path A, the two interfaces in the CC stacking are perfectly symmetric, which the ferroelectric polarizations at the two interfaces are equal and opposite. This means that the intermediate state is a typical antiferroelectric state. The polarization switching process in the GaSe-VPSe$_3$-GaSe sandwiched structure consequently involves a transition from ferroelectric CB stacking of 0.13 pC/m to an intermediate antiferroelectric CC stacking and then to a ferroelectric BC stacking of -0.13 pC/m, thereby completing the polarization reversal. Moreover, sliding ferroelectricity has also been experimentally reported in systems with three or more layers, which provides corroborating evidence for the experimental feasibility of our proposed scheme \cite{60,61}. According to DFT calculations, the intermediate antiferroelectric CC stacking is stabilized by 13.44 meV/f.u. compared to the ferroelectric phase.

\begin{figure*}[htb]
\begin{center}
\includegraphics[angle=0,width=1.0\linewidth]{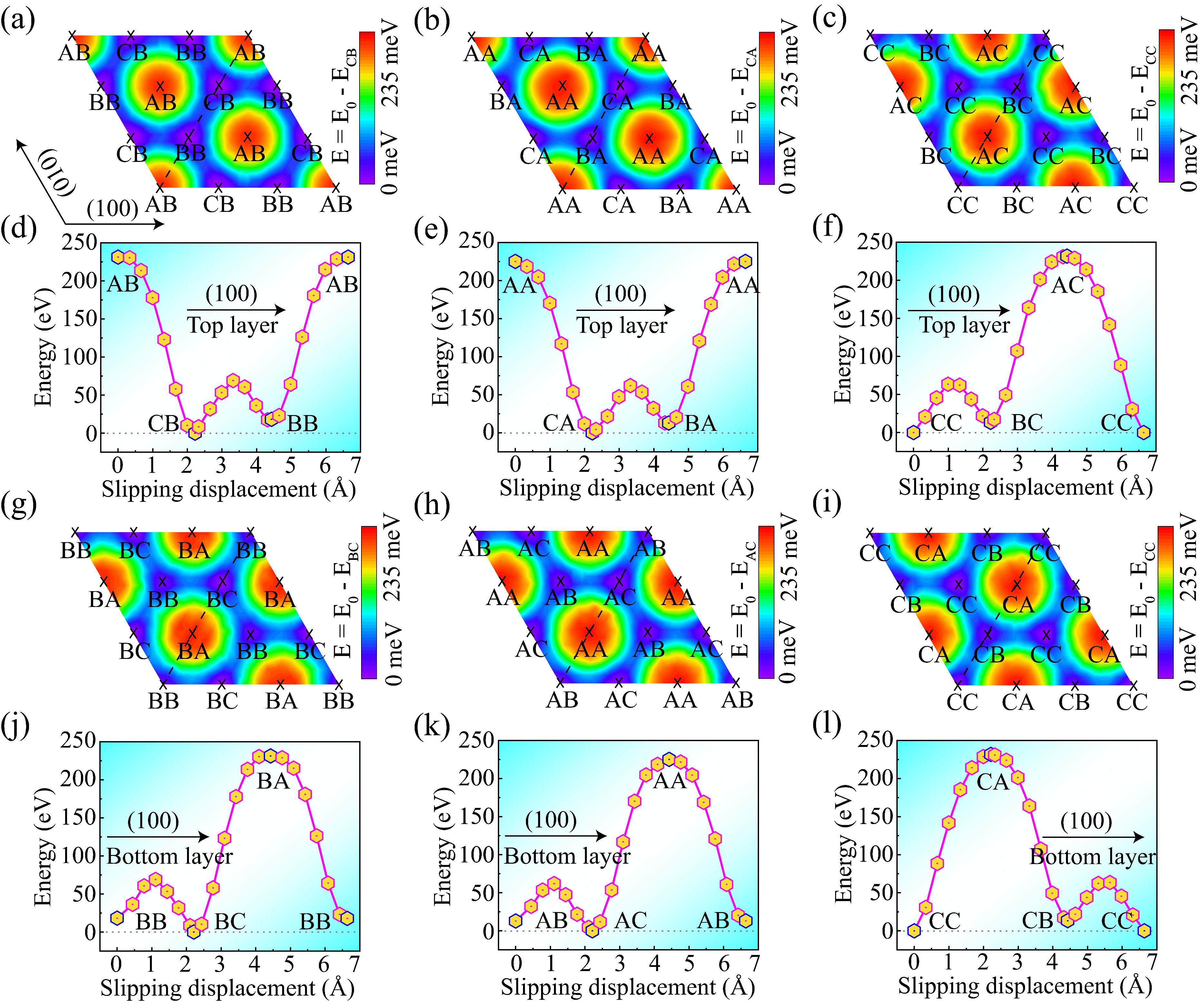}
\caption{
(a-c) The full space of top layer lateral shifts for (a) AB stacking, (b) AA stacking, and (c) CC stacking GaSe-VPSe$_3$-GaSe sandwiched structure. (d-f) Sliding energy barrier of top layer for (d) AB stacking, (e) AA stacking, and (f) CC stacking GaSe-VPSe$_3$-GaSe sandwiched structure along (100) direction. (g-i) The full space of bottom layer lateral shifts for (g) BB stacking, (h) AB stacking, and (i) CC stacking GaSe-VPSe$_3$-GaSe sandwiched structure. (j-l) Sliding energy barrier of bottom layer for (j) BB stacking, (k) AB stacking, and (l) CC stacking GaSe-VPSe$_3$-GaSe sandwiched structure along (100) direction.
}
\end{center}
\end{figure*}

To obtain a more comprehensive understanding of the optimal path of ferroelectric sliding in the GaSe-VPSe$_3$-GaSe sandwiched structure, as shown in Fig. 4(a-c, g-i), we systematically calculated the total energy of the kinds of XA, XB, XC, AX, BX, and CX stackings for the entire 2D space of lateral shifts. Here, X represents all possible stacking configurations. A 20 $\times$ 20 grid is adopted to utilize the lateral shifts. Fig. 4(a-c) depicts slip of the upper interface with a fixed lower interface, whereas Fig. 4(g-i) shows slid of the lower interface with a fixed upper interface. All possible slid paths are included in Fig. 4(a-c, g-i). To facilitate observation, the energy minima of all 2D spaces of lateral
shifts are normalized to zero. We can clearly observe that the stacking energies of AA, AB, BA, AC, and CA stackings reach the local maximum, while those of AB, BA, AC, CA, BB, BC, CB, and CC stackings show local minimum values. It is worth noting that AB, BA, AC, and CA stackings may be local maxima or local minima in different 2D spaces of lateral shifts.

In the following, let's comprehensively investigate the slip paths from AB stacking to BA stacking, CB stacking to BC stacking, and AC stacking to CA stacking. As previously categorized into two types, we can clearly observe that all slip paths from AB stacking to BA stacking and AC stacking to CA stacking involve transitions either from a local minimum to a local maximum, or from a local maximum to a local minimum. On the contrary, from stacking the CB sliding to the BC stacking, whether it is path A or path B, it is always from a local minimum to another local minimum. This indicates that among all possible paths, the sliding energy barrier from CB stacking to BC stacking is the lowest. Therefore, the transition path from the CB stacking to the BC stacking is the most optimal ferroelectric switching route in the GaSe-VPSe$_3$-GaSe sandwiched structure. Importantly, the CC stacking is the global minimum value (see Table SI). Consequently, the ferroelectric switching pathway that proceeds from CB stacking to BC stacking through the intermediate CC stacking is the most facile, exhibiting a minimal energy barrier of only 50.13 meV/f.u.. In addition, we calculate the sliding path along (100) for each 2D spaces. As shown in Fig. 4(d-f, j-l), they can also be divided into two categories. One type exhibits a double-well potential, as shown in Fig. 4(d, e), while the other type features a double-peak barrier, as depicted in Fig. 2(f, j-l). The energy barriers for both path A and path B of AB stacking to BA stacking, CB stacking to BC stacking, and AC stacking to CA stacking sliding transitions are all presented in Fig. 4(d-f, j-l).

\begin{figure*}[htb]
\begin{center}
\includegraphics[angle=0,width=1.0\linewidth]{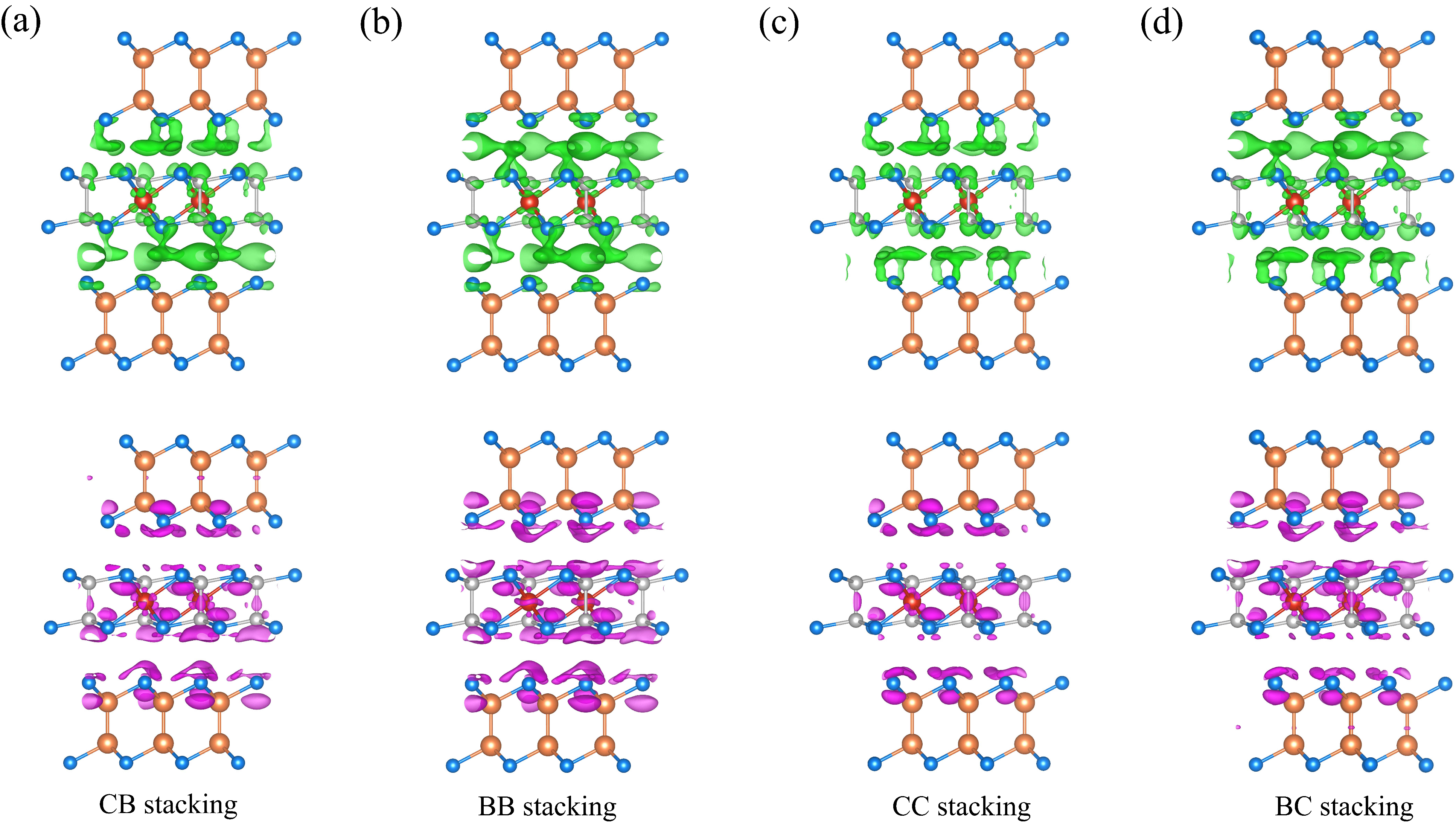}
\caption{
The differential charge density wave function of (a) CB stacking, (b) BB stacking, (c) CC stacking, and (d) BC stacking GaSe-VPSe$_3$-GaSe sandwiched structure. Green (upper) and magenta (lower) contours represent the gain and lose electrons, respectively.
}
\end{center}
\end{figure*}

To understand the microscopic mechanism behind the change in energy barriers from CB stacking to BC stacking, we employ the interlayer Hubbard model to describe it:
\begin{equation}
H = -\sum_{i,j,\sigma}t_{ij}(C_{i\sigma}^{\dag}C_{j\sigma} + H.c.) + \sum_{i}U_in_{i\uparrow}n_{i\downarrow},
\end{equation}
where the first term denotes hopping interaction and second term represent on-site Coulomb interaction. It is worth noting that the hopping and Coulomb interactions are positively correlated with the gain or loss of electrons. There are two paths for the transition from CB stacking to BC stacking: path A passes through CC stacking, while path B goes by BB stacking. When slipping from CB stacking to BB stacking, the Se-P distance has decreased from 4.26 $\rm \AA$ to 3.72 $\rm \AA$ at the upper layer. As shown in Fig. 5(a, b), this results in the simultaneous enhancement of both the gain and loss of electrons at the upper layer. Noted that the loss of electrons is greater than the gain of electrons, leading to a significant increase in Coulomb interactions. Therefore, the total energy of the system increased by 18.27 meV. On the contrary, when CB stacking slips to CC stacking, the Se-P distance in the lower interface has increased from 3.72 $\rm \AA$ to 4.26 $\rm \AA$. As shown in Fig. 5(a, c), the number of electrons gained and lost on the lower interface has decreased. The loss of electrons results in a greater reduction in the number of electrons gained, which means that the decrease in Coulomb repulsion is even more significant. Hence, the total energy of the system decreased by 13.44 meV. As shown in Fig. 5, the slip from either BB or CC stacking to BC stacking is exactly the reverse of the slip from CB stacking to either BB or CC stacking. Therefore, the most favorable slip pathway involves a transition from CB to CC stacking, followed by a subsequent transition to BC stacking.

\begin{figure*}[htb]
\begin{center}
\includegraphics[angle=0,width=1.0\linewidth]{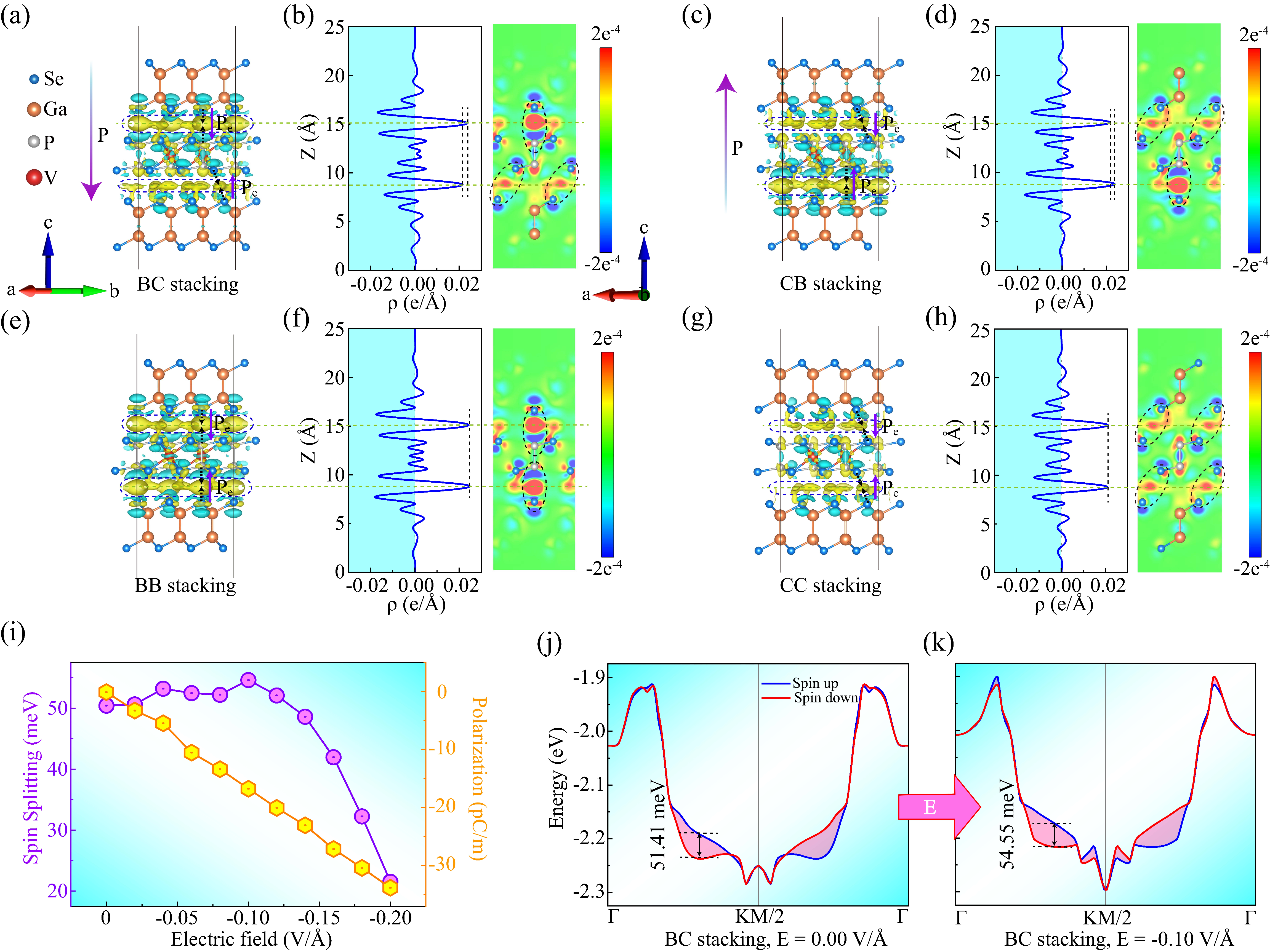}
\caption{
(a-h) 3D, planar averaged, and 2D differential charge density of (a, b) BC stacking, (c, d) CB stacking, (e, f) BB stacking, and (g, h) CC stacking GaSe-VPSe$_3$-GaSe sandwiched structure. (i) Spin splitting and ferroelectric polarization as functions of an external electric field for BC stacking. (j, k) Spin splitting of BC stacking at the (j) E = 0 V/${\rm \AA}$, and (k) E = -0.1 V/${\rm \AA}$ electric field.
}
\end{center}
\end{figure*}

\subsection{Mechanism Analysis and Electric Field Tuned Non-relativistic Spin Splitting}

Based on symmetry analysis, the sliding-induced ferroelectricity in AB, BA, BC, CB, AC, and CA stackings destroys the $\emph{PT}$ symmetry of VPSe$_3$. Consequently, a magnetic phase transition from the AFM to altermagnetism is realized. To uncover this mechanism, we investigated the charge transfer in GaSe-VPSe$_3$-GaSe sandwiched structure. Here, we take the optimal path from CB stacking to BC stacking as an example for an in-depth analysis. In the BC stacking, as shown in Fig. 6(a), two primary reasons are responsible for the downward ferroelectric polarization. The first part of the contribution comes from the upper interface Se-P atomic pairs. It does not involve a conventional sense of interatomic charge transfer \cite{62}. Instead, both Se and P lose electrons, leading to the formation of covalent bonds between the layers. It can be clearly observed that P loses more electrons than Se, thus contributing to the downward ferroelectric polarization. This phenomenon vanishes at the lower interface due to the misalignment of the Se-P atomic pair along the z-direction. The second part of the contribution comes from the Se-Se atomic pairs in the lower interface. Since the electronegativities of Se and Se atoms are the same, they cannot form a charge transfer between them. They also lost electrons, forming mushroom-shaped covalent bonds between the layers. This covalent bond formed between layers is reported by Ji $\emph{et al.}$ in their study of bilayer CrSe$_2$ \cite{63}. The resultant dipole moment points along the +z direction. Ultimately, the net dipole moment is -0.13 pC/m. To further prove this point, we investigate the xy-plane differential charge integrals for the BC stacking. As shown in Fig. 6(b), the formation of two covalent bonds is visualized at the upper and lower interfaces. The two interfacial covalent bonds generate a dipole moment along the -z axis. The mechanism for polarization reversal involves an external-electric-field-driven slide of GaSe relative to VPSe$_3$, which alters the relative positions of the Se-P and Se-Se atom pairs. In the CB stacking, as shown in Fig. 6(c, d), the situation is completely reversed from that of the BC stacking, resulting in 0.13 pC/m ferroelectric polarization along the +z direction.

Moreover, along the two transition pathways from CB stacking to BC stacking, the system passes through two intermediate states: BB and CC stackings. For the intermediate BB stacking, as shown in Fig. 6(e, f), the Se-P atomic pairs at the upper and lower interfaces are perfectly vertically aligned. The simultaneous loss of electrons from both Se and P leads to the formation of interlayer covalent bonds at the upper and lower interfaces. In the CC stacking, as shown in Fig. 6(g, h), the misalignment of Se-Se atoms in the upper and lower interfaces weakens the interlayer covalent bonding, which can significantly reduce the total energy of the system, making the CC stacking the most stable configuration. It can be clearly observed that the dipole moments generated by the covalent bonds at the upper and lower interfaces are arranged antiparallel to each other. It forms a typical antiferroelectricity. Therefore, the $\emph{PT}$ symmetry of the VPSe$_3$ has been restored. As a result, the two opposite spin sublattices can still be mutually mapped by inversion symmetry, thus preserving spin-simplified AFM in the sandwiched structure that is consistent with the monolayer VPSe$_3$, as shown in Fig. 2(g) and Fig. S3(d, g). The interface charge transfer in other stacking configurations is similar, as shown in Fig. S7.

Since the $\emph{PT}$ symmetry is broken by the out-of-plane ferroelectric polarization formed by the Se-P and Se-Se atomic pair is the primary origin of altermagnetism, any type of out-of-plane polarization can be utilized for this purpose. An external electric field precisely meets this requirement and is also the most straightforward manipulation method in experiments. Fig. 6(i) exhibits that the electric field tunes the magnitude of spin splitting and ferroelectric polarization for the BC stacking. A clear observation is that the ferroelectric polarization increases linearly with the external electric field, while the spin splitting first increases slowly and then decreases rapidly. When the external electric field increases to -0.10 V/$\rm \AA$, as shown in Fig. 6(j, k), the spin splitting and ferroelectric polarization increases from 51.41 meV and -0.13 pC/m to 54.55 meV and -16.77 pC/m, respectively. The NRSS for typical stacking configurations is listed in Table SIII. More interestingly, as shown in Fig. S8, the band structures undergo a rich evolution along the $\Gamma$-KM/2-$\Gamma$ path at $\sim$E$_F$ - 2.3 eV. In the absence of an electric field, the spins is degenerate near the KM/2 point, while an external electric field lifts this degeneracy. The phenomenon is most obvious at -0.08 V/$\rm \AA$ (see Fig. S8(d)). Moreover, the external electric field can increase the number of van Hove singularity. This indicates that not only can the altermagnetic phase be manipulated by sliding ferroelectric, but also its van Hove singularities and spin splitting can be further tuned through an external electric field.

\section{CONCLUSION}
In conclusion, we present a novel sliding ferroelectric in multilayer heterostructure switchable altermagnetism effect: the ferroelectric polarization reversal effectively switches the altermagnetic spin-splitting. Based on symmetry analysis and DFT calculations, we demonstrate the mechanism in the GaSe-VPSe$_3$-GaSe sandwiched structure. Interlayer sliding can break and restore the $\emph{PT}$ symmetry, which enables a switching between altermagnetic and conventional AFM states along with an effective modulation of the NRSS. We have systematically investigated all the transition pathways in the GaSe-VPSe$_3$-GaSe sandwiched structure. Our calculations identify the CB$\rightarrow$CC$\rightarrow$BC stackings pathway as the most favorable, with a mere 50.13 meV/f.u. energy barrier, connecting the initial and final ferroelectric states via an intermediate antiferroelectric phase. Interestingly, the origin of the magnetic phase transition is not the interfacial charge transfer as previously thought, but rather the formation of interlayer covalent bonds between Se-Se or Se-P atomic pairs. In addition, the magnitude of NRSS and ferroelectric polarization can be effectively tuned by an external electric field. Our work provides a demonstration of the tight coupling between altermagnetic spin splitting and sliding ferroelectric, making it an ideal platform for electric field controllable altermagnetic multiferroic devices.

\section*{Keywords}
Altermagnetic multiferroic, Sliding ferroelectricity, Strong magnetoelectric effect, Magnetic phase transition, Interfacial covalent bond

\section*{CRediT authorship contribution statement}
Pengqiang Dong: Methodology, Validation, Investigation, Software, Data curation, Visualization. Hanbo Sun: Investigation. Chao Wu: Investigation. Ping Li: Conceptualization,  Supervision,
Formal analysis, Resources, Writing-original draft, Project administration, Funding acquisition.

\section*{Acknowledgements}
This work is supported by the National Natural Science Foundation of China (Grants No. 12474238, and No. 12004295), P. Li also acknowledge supports from the Shaanxi Youth Science and Technology New Star Project (Grant No. 2025ZC-KJXX-71), the China's Postdoctoral Science Foundation funded project (No. 2022M722547), the Fundamental Research Funds for the Central Universities (xzy012025031), the Open Project of National Laboratory of Solid State Microstructures (No. M38035), and the Open Project of State Key Laboratory of Silicon and Advanced Semiconductor Materials (No. SKL2024-10).

\section*{Supplementary materials}
Supplementary material associated with this article can be found, in the online version. at doi:xxx.xxx.


\end{document}